\documentstyle[12pt]{article}
\begin{document}

\author{A. de Souza Dutra\thanks{%
e-mail: dutra@feg.unesp.br} \\
%EndAName
UNESP/Campus de Guaratinguet\'a-DFQ\\
Av. Dr. Ariberto Pereira da Cunha, 333\\
Guaratinguet\'a - SP - Brasil\\
Cep 12516-410 \and C. A. S. Almeida \\
%EndAName
Departamento de F\'{\i}sica, Universidade Federal do Cear\'a\\
Centro de Ci\^encias, Campus do Pici, Caixa Postal 6030\\
CEP 60455-760, Fortaleza - Cear\'a - Brasil}
\title{{\Large {Exact solvability of potentials with spatially dependent effective
masses} }}
\maketitle

\begin{abstract}
We discuss the relationship between exact solvability of the Schroedinger
equation, due to a spatially dependent mass, and the ordering ambiguity.
Some examples show that, even in this case, one can find exact solutions.
Furthermore, it is demonstrated that operators with linear dependence on the
momentum are nonambiguous.
\end{abstract}

\newpage

\section{Introduction\ }

Along the years, many people has studied the Schroedinger equation regarding
its exact solvability. This because, despite of the intrinsic interest of
the systems exactly solved, these solutions can be used to get better
approximated solutions for potentials more physically interesting. In fact
in the last decades many advances were put forward in this area by doing the
classification of the quantum potentials regarding its solvability, for
instance by relating the solutions to an underlying supersymmetry \cite
{avnash}, or a dynamical one \cite{henrique}. Similar discussions are done
for the case of the so called quasi exactly solvable potentials \cite{dutra1}%
, and the conditionally exactly solvable ones \cite{ces}. In particular the
exact potentials are in general considered for constant or at most
time-dependent masses \cite{mostafadeh}-\cite{abdalla}. Here we intend to
show that when we take into account a spatial dependence of the mass, there
are other potentials that are not solvable when the masses are constant.
Furthermore, these potentials will lead to eigenfunctions and eigenvalues
depending on the chosen ordering of the Hamiltonian operator.

On the other hand, the problem of ordering ambiguity is a long standing one
in quantum mechanics. Some of the founders of quantum mechanics as Born and
Jordan, Weyl, Dirac and von Newmann did work in this matter, see for
instance the excellent critical review by Shewell \cite{shewell}. There are
many examples of physically important systems, for which such ambiguity is
quite relevant. For instance we can cite the problem of impurities in
crystals \cite{luttinger}\cite{wannier}\cite{slater}, the dependence of
nuclear forces on the relative velocity of the two nucleons \cite{rojo}\cite
{razavy}, and more recently the study of semiconductor heterostructures \cite
{bastard}\cite{weisbuch}. A very important example of ordering ambiguity is
that of the minimal coupling in systems of charged particles interacting
with magnetic fields \cite{landau}. This problem has as the accepted
solution, the ordering of the ambiguous term $\vec{A}(\vec{x}).\vec{p}$ by
defining a symmetrized one (Weyl ordering) $(\vec{A}(\vec{x}).\vec{p}\,+\,%
\vec{p}.\vec{A}(\vec{x}))/2$. By the way this is even used to define the
prescription to evaluate the Feynmann path integral at the mean point \cite
{feynman}\cite{schulman}\cite{simao}. The way one chooses the point to
evaluate the path integral is shown to be closely related to the problem of
ordering ambiguity \cite{kerner}\cite{cohen}. In fact the so called Weyl
ordering is usually accepted by some textbooks as the correct one \cite
{tdlee}\cite{gasiorowicz}. However we also show that, no matter the ordering
used to the term $\vec{A}(\vec{x}).\vec{p}$, the same result is obtained,
{\it i. e. }there is no ambiguity.

Notwithstanding, taking into account the spatial variation of the
semiconductor type, some effective Hamiltonians are proposed with a
spatially dependent mass for the carrier \cite{duke}-\cite{carlos}. In this
last work, it was tried to circumvent the problem of ambiguity, by starting
from the Dirac equation which does not have any ambiguity, and then taking
the nonrelativistic limit. In this process the authors advocated in favor of
the Li and Kuhn \cite{li} proposal of effective Hamiltonian. One of the
goals of this work is to show that the Hamiltonian proposed by Li and Kuhn
\cite{li}\cite{carlos}, is in fact equivalent to that coming from the Weyl
ordering.

We also suggest that these exact solutions could be used as a kind of guide
to, at least, restrict the possible choices of ordering. The principal idea
is to suppose that, once one have found the ordering without ambiguity for a
given potential or class of potentials, that ordering should be extended to
remaining physical potentials.

\section{Generalized Effective Schroedinger Equation}

We start this section by defining a quite general Hermitian effective
Hamiltonian for the case of a spatially dependent mass. In general it is
used the Hamiltonian proposed by von Roos \cite{vonroos},
\begin{equation}
H_{VR}\,=\,\frac{1}{4}\left[ m^{\alpha }(\vec{r})\,\hat{p}\,m^{\beta }(\vec{r%
})\,\hat{p}\,m^{\gamma }(\vec{r})\,\,+\,m^{\gamma }(\vec{r})\,\hat{p}%
\,m^{\beta }(\vec{r})\,\hat{p}\,m^{\alpha }(\vec{r})\,\right] ,
\end{equation}

\noindent but to accommodate the possibility of including the case of the
Weyl ordering \cite{simao} in a more evident way, we will use an effective
Hamiltonian with four terms given by
\[
H\,=\,\frac 1{4(a+1)}\left\{ a\left[ m^{-1}(\vec r)\,\hat p^2\,+\,\,\hat
p^2\,m^{-1}(\vec r)\right] \,+\,m^\alpha (\vec r)\,\hat p\,m^\beta (\vec
r)\,\hat p\,m^\gamma (\vec r)\,\,+\,\right. \,
\]
\begin{equation}
\left. +\,\,\,m^\gamma (\vec r)\,\hat p\,m^\beta (\vec r)\,\hat p\,m^\alpha
(\vec r)\right\} .
\end{equation}

\noindent In both cases there is a constraint over the parameters: $\alpha
+\beta +\gamma =-1$, and the Weyl ordering is recovered when one choose $%
a=1,\,\alpha \,=\,\gamma \,=\,0$. A similar Hamiltonian was used by Levinger
and collaborators \cite{rojo}\cite{razavy}.

Using the properties of the canonical commutators, it is easy to show that
one can put the momenta to the right, so obtaining the following effective
Hamiltonian
\begin{equation}
H\,=\,\frac{1}{2\,m}\,\hat{p}^{2}\,+\,\frac{i\hbar }{2}\frac{dm/dr}{m^{2}}%
\,\,\hat{p}\,\,+\,U_{\alpha \beta \gamma a}\left( r\right) ,
\end{equation}

\noindent where
\[
U_{\alpha \beta \gamma a}\left( r\right) \,=\,-\frac{\hbar ^{2}}{%
4m^{3}\left( a+1\right) }\left[ \left( \alpha +\gamma -a\right) m\,\frac{%
d^{2}m}{dr^{2}}\,+\,\right.
\]
\begin{equation}
\left. +\,\,\,2\,\left( \,a\,-\alpha \,\gamma \,-\,\alpha \,-\,\gamma
\right) \left( \frac{dm}{dr}\right) ^{2}\right] .
\end{equation}

It is curious to note that all the ambiguity is in the last effective
potential term, and that it can be eliminated by imposing some convenient
constraints over the ambiguity parameters, namely
\begin{equation}
\alpha \,+\,\gamma \,-\,a\,=\,0,\,\,\,\,a\,-\alpha \,\gamma \,-\,\alpha
\,-\,\gamma =\,0,
\end{equation}

\noindent which have two equivalent solutions, (i) $\alpha =0$ and $a=\gamma
$, or (ii) $a=\alpha $ and $\gamma =0$. In this case the effective
Schroedinger equation will not depend on the ambiguity parameters, but will
contain a first order derivative term. In the next section, we will be
interested in getting exact solutions of the resulting equation for some
particular potentials, and trying to get some information about the proposed
orderings appearing in the literature.

\section{Exact Potentials with Coordinate Dependent Mass}

First of all, we will rewrite the Schroedinger equation coming from the
above effective Hamiltonian
\begin{equation}
-\frac{\hbar ^2}{2m}\frac{d^2\psi }{dr^2}\,+\,\frac{\hbar ^2}2\left( \frac{%
dm/dr}{m^2}\right) \frac{d\psi }{dr}\,+\,\left( V\left( r\right)
\,+\,U_{\alpha \beta \gamma a}\left( r\right) \,-\,E\right) \psi \,=\,0,
\end{equation}

\noindent where we introduce, for the sake of generality, a potential term $%
V\left( r\right) $. It is not difficult to verify that doing a wave function
redefinition
\[
\psi \left( r\right) \,=\,m^{\frac{1}{2}}\,\varphi \left( r\right) ,
\]

\noindent one get a differential equation in a more familiar form
\begin{equation}
-\frac{\hbar ^2}{2m}\frac{d^2\varphi }{dr^2}\,+\,\left( U_{eff}\,-\,E\right)
\varphi \left( r\right) \,=\,0,
\end{equation}

\noindent with the effective potential defined through
\begin{equation}
U_{eff}\,=\,V\left( r\right) \,+\,U_{\alpha \beta \gamma a}\left( r\right)
\,+\,\frac{\hbar ^{2}}{4m}\left( \frac{3}{2}\left( \frac{dm/dr}{m}\right)
^{2}\,-\,\frac{d^{2}m/dr^{2}}{m}\right) .  \label{ref0}
\end{equation}

At this point it is interesting to discuss about the exact solvability of a
given system. From above equations, and noting that one could recover an
usual Schroedinger equation with a constant unitary mass by multiplying it
by $m(x)$. One can easily verify that whenever $m(x)\,\left( U_{eff}\left(
x\right) \,-\,E\right) =\,V_{N}\left( x\right) \,-\,{\cal E}$, where $%
V_{N}\left( x\right) $ is an exact \cite{avnash}, quasi-exact \cite{henrique}
or conditionally exact \cite{ces} potential, and besides ${\cal E}$ is a
constant term, we will have the original pair of potential $V\left( x\right)
$ and $m\left( x\right) $, with this same kind of solvability property.

Now let us take some particular cases where one have exact solution for the
above equation. As the first example we consider a particle with
exponentially decaying or increasing mass, in the presence of a potential
with similar behavior,
\begin{equation}
m\left( x\right) \,=\,m_{0}\,e^{c\,x},\,\,V\left( x\right)
\,=\,V_{0\,}\,e^{c\,x},
\end{equation}

\noindent and in this case we will have as the effective equation
\begin{equation}
-\frac{\hbar ^{2}}{2m_{0}}\frac{d^{2}\varphi }{dx^{2}}\,+\,\,\left(
_{\,}V_{0}\,e^{2\,c\,x}\,-\,E\,e^{c\,x}\right) \,\varphi \,=\,{\cal E}%
\,\varphi ,\,  \label{ref1}
\end{equation}

\noindent where ${\cal E}\,\equiv \frac{\hbar ^{2}}{m_{0}}\left( q\,-\,\frac{%
c^{2}}{8}\right) $, and we defined $q$ as
\begin{equation}
q\,=\,\frac{c^{2}}{4\left( a+1\right) }\,\left( a\,-\,2\,\alpha \,\gamma
\,-\,\alpha \,-\,\gamma \right) .
\end{equation}

Note that this last equation corresponds to a Schroedinger equation for a
particle with constant mass under the influence of the Morse potential \cite
{morse}. However we will make a further modification that is convenient to
put it in a more familiar form, that of a harmonic oscillator with
centripetal barrier. This is done by using a coordinate transformation and a
wave function redefinition defined as \cite{ces}
\begin{equation}
x\,=\,f\left( u\right) ,\,\varphi \left( x\right) \,=\,\sqrt{\frac{df\left(
u\right) }{du}\,}\,\chi \left( u\right) ,
\end{equation}

\noindent so that, after straightforward calculations one ends with
\begin{equation}
-\frac{\hbar ^{2}}{2m_{0}}\frac{d^{2}\chi }{du^{2}}\,+\,\,\left( W_{T}\left(
u\right) \,-\,E_{T}\right) \,\chi \,=0,  \label{ref2}
\end{equation}

\noindent with the following definitions
\begin{equation}  \label{ref3}
W_T\left( u\right) \,-\,E_T\,=\,\left( \frac{df\left( u\right) }{du}\right)
^2\left[ W\left( u\right) \,-\,E\right] \,-\,\frac{\hbar ^2}4\left[ \frac{%
d^3f\left( u\right) /du^3}{df\left( u\right) /du}\,-\,\frac 32\left( \frac{%
d^2f\left( u\right) /du^2}{df\left( u\right) /du}\right) ^2\right] ,
\end{equation}

\noindent and it was used that in Eq. \ref{ref1}, $W\left( u\right)
\,=\,m_{0\,}V_{0}\,e^{2\,c\,f\left( u\right) }\,-\,E\,e^{c\,f\left( u\right)
}$. In the present case $f\left( u\right) \,=\,\ln \left( u^{\frac{2}{c}%
}\right) $and this lead us to
\[
-\frac{\hbar ^{2}}{2}\frac{d^{2}\chi }{du^{2}}\,+\,\,\left( \frac{\omega ^{2}%
}{2}\,u^{2}\,-\,\frac{g}{u^{2}}\right) \,\chi \,\,=\,E_{T}\,\chi ,\,\omega
\,=\,\frac{\sqrt{2\,m_{0\,}V_{0}}}{c},\,
\]
\begin{equation}
g\,=\,\frac{{\cal E}}{c^{2}}\,+\,\frac{\hbar ^{2}}{8},\,E_{T}\,=\,\frac{%
m_{0}\,E}{c^{2}},\,
\end{equation}

\noindent From which one can easily reconstruct the wave function of the
original potential and the respective energy spectrum. This last is given by
\begin{equation}
E_{n}\,=\,\left( \hbar \,c\right) \sqrt{\frac{2\,V_{0}}{m_{0}}}\,\left[
2\,n\,+1\,+\,\nu \left( \alpha ,\beta ,\gamma ,a\right) \right] ,\,\nu
\left( \alpha ,\beta ,\gamma ,a\right) \,=\,\left( \frac{1}{4}\,-\,\frac{2\,q%
}{c^{2}}\right) .
\end{equation}

Now we study the effect of using some of the orderings appearing in the
literature. By using the Gora and Williams ordering (a$=\gamma \,=0,\,\alpha
\,=\,-1$) \cite{gora} or that due to BenDaniel and Duke\thinspace ($%
a\,=\,\alpha \,=\,\gamma \,=0$) \cite{duke}, one ends with a complex energy
because in these cases one have $\nu \left( \alpha ,\beta ,\gamma ,a\right)
\,=\,\frac i2$. If one consider that the good ordering should lead to
sensible results for any potential, these orderings should be discarded. On
the other hand, once what is observable are the differences between the
energy levels, it is easy to conclude that in this case the spectrum is
nonambiguous. Furthermore for the orderings of Zhu and Kroemer \cite{zhu} ($%
a=0,\,\alpha =-\frac 12\,=\,\gamma $), Li and Kuhn \cite{li} ($a=0,\,\alpha
=0,\,\gamma =-\frac 12$), and Weyl \cite{tdlee} ($a=1,\,\alpha =0=\gamma $),
the ambiguous term $\nu $ is zero.

The next example is that of a quadratically growing mass in the presence of
a singular potential
\begin{equation}
m\left( x\right) \,=\,c\,x^2,\,\,V\left( x\right) \,=\,\frac
A{c\,x^4}\,+\,\frac B{c\,x^2}.
\end{equation}

\noindent In this case the corresponding effective equation, analogously to
the equation (\ref{ref1}) is given by
\begin{equation}
-\frac{\hbar ^{2}}{2}\frac{d^{2}\varphi }{dx^{2}}\,+\,\,\left( -\,\left(
c\,E\right) \,x^{2}\,+\,\frac{\left( A\,+\,g\right) }{x^{2}}\right)
\,\varphi \,=\,-B\,\varphi ,
\end{equation}

\noindent where $g$ comes from the ambiguity correction and is given by
\[
g=\frac{\hbar ^{2}}{2\left( a+1\right) }\left[ \,4\,\alpha \,\gamma
\,+\,3\left( \alpha \,+\,\gamma \right) \,-\,a\,+\,2\right] .
\]

\noindent Here, using once again the solution for the harmonic oscillator
with centripetal barrier, the energy spectrum is written as
\begin{equation}
E_{n}\,=\,-\,\frac{2\,B^{2}}{c\,\left( 2\,n\,+\nu \,+\,1\right) ^{2}\,\hbar
^{2}},\,\,\nu (\alpha ,\beta ,\gamma ,a,A)\,=\,\left[ \frac{1}{4}\,-\,\frac{%
2\left( A\,+\,g\right) }{\hbar ^{2}}\right] ^{\frac{1}{2}}.
\end{equation}

\noindent Now it is very clear that the ambiguity can not be avoided through
some redefinition of the zero point energy. Let us now to evaluate the
results coming from the cases proposed in the literature, for Gora and
Williams case the function $\nu \,=\,\left[ \frac{5}{4}\,-\,\frac{2\,A}{%
\hbar ^{2}}\right] ^{\frac{1}{2}}$, for that due to BenDaniel and Duke
\thinspace $\nu \,=\,\left[ -\frac{7}{4}\,-\,\frac{2\,A}{\hbar ^{2}}\right]
^{\frac{1}{2}}$, Zhu and Kroemer gives $\nu \,=\,\left[ \frac{1}{4}\,-\,%
\frac{2\,A}{\hbar ^{2}}\right] ^{\frac{1}{2}}$, for Li and Kuhn $\nu
\,=\,\left[ -\frac{1}{4}\,-\,\frac{2\,A}{\hbar ^{2}}\right] ^{\frac{1}{2}}$,
and when using Weyl ordering one gets again the same result of Li and Kuhn.
By the way it is interesting to note this feature and in fact we will show
that these orderings are equivalent in the next section.

\section{Equivalence of Weyl and Li and Kuhn orderings}

As can be easily checked from the equation (\ref{ref0}), the effective
potentials of both these orderings are equal,
\begin{equation}
U_{eff}\,=\,V\left( r\right) \,\,+\,\frac{\hbar ^{2}}{8\,m}\left( \left(
\frac{dm/dr}{m}\right) ^{2}\,-\,\frac{d^{2}m/dr^{2}}{m}\right) ,
\end{equation}

\noindent but we will demonstrate the equivalence by starting with the Weyl
ordered Hamiltonian
\begin{equation}
H_{Weyl}\,=\,\frac 18\left[ \frac 1m\,\hat p^2\,+\,\,\hat p^2\frac
1m\,+\,2\,\,\hat p\,\frac 1m\,\,\,\hat p\right] ,
\end{equation}

\noindent and using that one can rewritten the operators suitably,
\[
\,\hat p\frac 1m\,\,\hat p\,=\,\frac 1{\sqrt{m}}\,\,\hat p\,\frac 1{\sqrt{m}%
}\,\,\hat p\,+\,\,\hat p\,\frac 1{\sqrt{m}}\,\,\hat p\,\frac 1{\sqrt{m}%
}\,+\, \frac{i\,\hbar }2\left[ \left( \frac{dm/dr}{m^2}\right) \,\,\hat
p\,-\,\,\hat p\,\left( \frac{dm/dr}{m^2}\right) \right] ,
\]
\[
\,\hat p^2\,\frac 1m\,=\,\,\hat p\,\frac 1{\sqrt{m}}\,\,\hat p\,\frac 1{%
\sqrt{m}}\,+\,\frac{i\,\hbar }2\,\hat p\,\left( \frac{dm/dr}{m^2}\right) ,
\]
\begin{equation}
\frac 1m\,\hat p^2\,=\,\frac 1{\sqrt{m}}\,\,\hat p\,\frac 1{\sqrt{m}%
}\,\,\hat p\,-\,\frac{i\,\hbar }2\,\left( \frac{dm/dr}{m^2}\right) \,\,\hat
p\,,
\end{equation}

\noindent substituting these operators in the Weyl Hamiltonian, it becomes
equal to
\begin{equation}
H_{Weyl}\,=\,\frac 14\left[ \frac 1{\sqrt{m}}\,\,\hat p\,\frac 1{\sqrt{m}%
}\,\,\hat p\,+\,\,\hat p\,\frac 1{\sqrt{m}}\,\,\hat p\,\frac 1{\sqrt{m}%
}\right] \,=\,H_{LK},
\end{equation}

\noindent which is precisely the Hamiltonian proposed by Li and Kuhn \cite
{li}. This ends our demonstration of the equivalence of these two orderings.
Now it is important to stress that this implies that this ordering has the
advantage being that which comes naturally from the corresponding
nonambiguous Dirac equation \cite{carlos}.

In the next, we show that when the ambiguous term is linear in the momentum
(classically $H_{clas}=f\left( x\right) \,p$), contrary to what is usually
believed, there is no ambiguity in fact. For this we observe that
\begin{equation}
f\left( x\right) ^{\alpha }\,p\,f\left( x\right) ^{\beta }\,=\,f\left(
x\right) \,p\,-\,i\,\hbar \,\beta \,\frac{df}{dx},\,\,\alpha \,+\,\beta
\,=\,1,
\end{equation}
\begin{equation}
\frac{\left( f\left( x\right) ^{\alpha }\,p\,f\left( x\right) ^{\beta
}\,+\,f\left( x\right) ^{\beta }\,p\,f\left( x\right) ^{\alpha }\right) }{2}%
\,=\,f\left( x\right) \,p\,-\,i\,\hbar \,\frac{df}{dx}.
\end{equation}

\noindent So, any Hermitian construction of the quantum Hamiltonian will be
necessarily equivalent to that due to any other, and consequently
nonambiguous. At this point it is important to remark that the common notion
that, the so called minimum coupling principle for the introduction of the
electromagnetic interaction implies into the Weyl ordering as the privileged
one, is really misleading once the above demonstration lead us to conclude
that any ordering will conduce essentially to the same Schroedinger equation
in this case.

\section{Conclusions}

In this work we have discussed the problem of solvability and ordering
ambiguity in quantum mechanics. It was shown through particular examples
that the exact solvability depends not only on the form of the potential,
but also on the spatial dependence on the mass. In the first example we have
considered an exponentially decaying or increasing mass and potential, which
was mapped into a harmonic oscillator with centripetal barrier. In this case
the energy levels could be redefined in such a way that the ordering
ambiguity disappears. In the second case however, a quadratically growing
mass and a singular potential, this can not be done. Furthermore we did
observe that some orderings proposed in the literature lead us to non
physically acceptable energies, and could possibly be discarded. In both
cases we did note that the orderings due Weyl and Li and Kuhn lead us to the
same results. This stimulated us to demonstrate that, in fact, they are
equivalent. In the second example we did perceive that some kind of extremum
principle could possibly be used to render the problem less ambiguous.

Finally we demonstrate that operators with linear dependence on the momentum
are totally nonambiguous, leading to the conclusion that there is no reason,
from this point of view, to choose the mean point expansion in the path
integral formalism, as is usually accepted \cite{feynman}\cite{schulman}. As
far as we know, this feature was not perceived until now.

\medskip

\noindent {\bf Acknowledgments:} The authors are grateful to CNPq and FAPESP
for partial financial support. The authors also acknowledge M. Hott for a
careful reading of the manuscript.

\end{document}